\documentclass[conference,onecolumn]{IEEEtran}
\usepackage{cite}

\usepackage{subfigure}  % use for side-by-side figures
\usepackage[pdftex]{graphicx}   % need for figures

\usepackage[cmex10]{amsmath}
\newcommand{\eqs}{\begin{eqnarray*}}
\newcommand{\eqf}{\end{eqnarray*}}
\newcommand{\lef}{\left(}
\newcommand{\rig}{\right)}
\newcommand{\mas}{\begin{array}}
\newcommand{\maf}{\end{array}}
\newcommand{\deriv}{\textnormal{d}}

\begin{document}

\title{Electron relaxation times and resistivity in metallic nanowires due to tilted grain boundary planes}

\author{\IEEEauthorblockN{Kristof Moors\IEEEauthorrefmark{1}\IEEEauthorrefmark{2},
Bart Sor\'ee\IEEEauthorrefmark{1}\IEEEauthorrefmark{3}\IEEEauthorrefmark{4} and
Wim Magnus\IEEEauthorrefmark{1}\IEEEauthorrefmark{3}}
\IEEEauthorblockA{\IEEEauthorrefmark{1}imec, Kapeldreef 75, B-3001 Leuven, Belgium}
\IEEEauthorblockA{\IEEEauthorrefmark{2}Institute for Theoretical Physics, KU Leuven, Celestijnenlaan 200D, B-3001 Leuven, Belgium}
\IEEEauthorblockA{\IEEEauthorrefmark{3}Physics Department, University of Antwerp, Groenenborgerlaan 171, B-2020 Antwerpen, Belgium}
\IEEEauthorblockA{\IEEEauthorrefmark{4}Electrical Engineering (ESAT) Department, KU Leuven, Kasteelpark Arenberg 10, B-3001 Leuven, Belgium\\Email: kristof@itf.fys.kuleuven.be}}

\maketitle

\begin{abstract}
%\boldmath
We calculate the resistivity contribution of tilted grain boundaries with varying parameters in sub-10nm diameter metal nanowires. The results have been obtained with the Boltzmann transport equation and Fermi's golden rule, retrieving correct state-dependent relaxation times. The standard approximation schemes for the relaxation times are shown to fail when grain boundary tilt is considered. Grain boundaries tilted under the same angle or randomly tilted induce a resistivity decrease.
\end{abstract}

\section{Introduction}

A widely used model to identify the resistivity contributions due to different electron scattering mechanisms in metal interconnects is the Mayadas-Shatzkes (MS) model \cite{mayadas1970electrical}. In addition to a bulk term, this model includes partially diffusive scattering at the wire boundary surface and scattering at grain boundaries (GBs), the two dominant scattering mechanisms in small diameter wires \cite{steinhogl2002size,guillaumond2003analysis,steinhogl2004comprehensive,chawla2011electron}. The MS model can be used to fit measured resistivity with different diameters: a certain diffusive boundary scattering probability, GB density and barrier height such that the relative contribution of surface scattering and GB scattering can be obtained. Regardless the simplicity of the model, it has many shortcomings that become worse when the wire diameter is reduced. First, the Fuchs-parameter is used and represents the probability for diffusive scattering at the wire surface \cite{fuchs1938conductivity,sondheimer1952mean}. This parameter is not directly based on physical properties of the boundaries, e.g. does not take into account the roughness of the wire surface and neglects the quantum-mechanical aspects of confined electron states. Secondly, the GBs are represented as potential Dirac delta barrier planes perpendicular to the transport direction, which limits the scattering possibilities to Umklapp scattering. The resistivity is expected to be different when other orientations are considered. Finally, the electron relaxation times (RTs) are obtained approximately, neglecting the coupling of different electron states and their RT.

In \cite{moors2014resistivity} was shown recently that the resistivity contribution of GBs in thin metal nanowires (below 10nm diameter) is much larger than the surface roughness contribution, while they are found to be almost equally responsible for the resistivity contribution in larger diameter nanowires. The correct state-dependent RTs are obtained and surface roughness scattering probabilities are retrieved from Ando's model \cite{ando1982electronic}, describing surface scattering quantum-mechanically with use of the roughness properties of the boundary surface \cite{moors2014resistivity}.

We extend this work here by considering tilted GB planes and obtain the correct electron RTs and corresponding resistivity. We consider two models for tilted GB planes: tilted GB planes having the same fixed angle and randomly tilted GB planes without correlations between them.

\section{Model}
The conduction electrons are modeled in the same way as in \cite{moors2014resistivity}, as particles with effective mass approximation (effective mass denoted by $m_e^*$) in a box with square cross-section (diameter $D$). The box is an infinite potential well with periodic boundary conditions in the transport direction, for which the wire dimension is much larger than the transverse dimensions.

The Boltzmann transport equation for a stationary state and the resulting expression for the current $J_z$ are given by:
\begin{align} \label{BTE}
-\frac{e E_z}{\hbar} \frac{\partial f_{n_x n_y}\lef k_z \rig}{\partial k_z} &= \left.\frac{\partial f_{n_x n_y}\lef k_z \rig}{\partial t}\right|_\textnormal{collisions} = - \frac{f_{n_x n_y} \lef k_z \rig - f^{\textnormal{eq.}}_{n_x n_y} \lef k_z \rig}{\tau_{n_x n_y}\lef k_z \rig}, \\ \notag
J_z &= -\frac{e}{\pi} \sum_{n_x, n_y} \int\limits_{-\infty}^{+\infty} \deriv k_z \frac{ \hbar k_z }{m^*_e} f_{n_x n_y} \lef k_z \rig,
\end{align}
with $E_z$ the electric field (along the transport direction), $k_z$ the wave vector along the transport direction, $n_{x/y}$ the sub-band indices for the transverse dimensions and $f_{n_x,n_y}\lef k_z \rig$ the distribution function for a specific sub-band.

\begin{figure}[htb]
\centering
\subfigure[]{\includegraphics[width=0.39\linewidth]{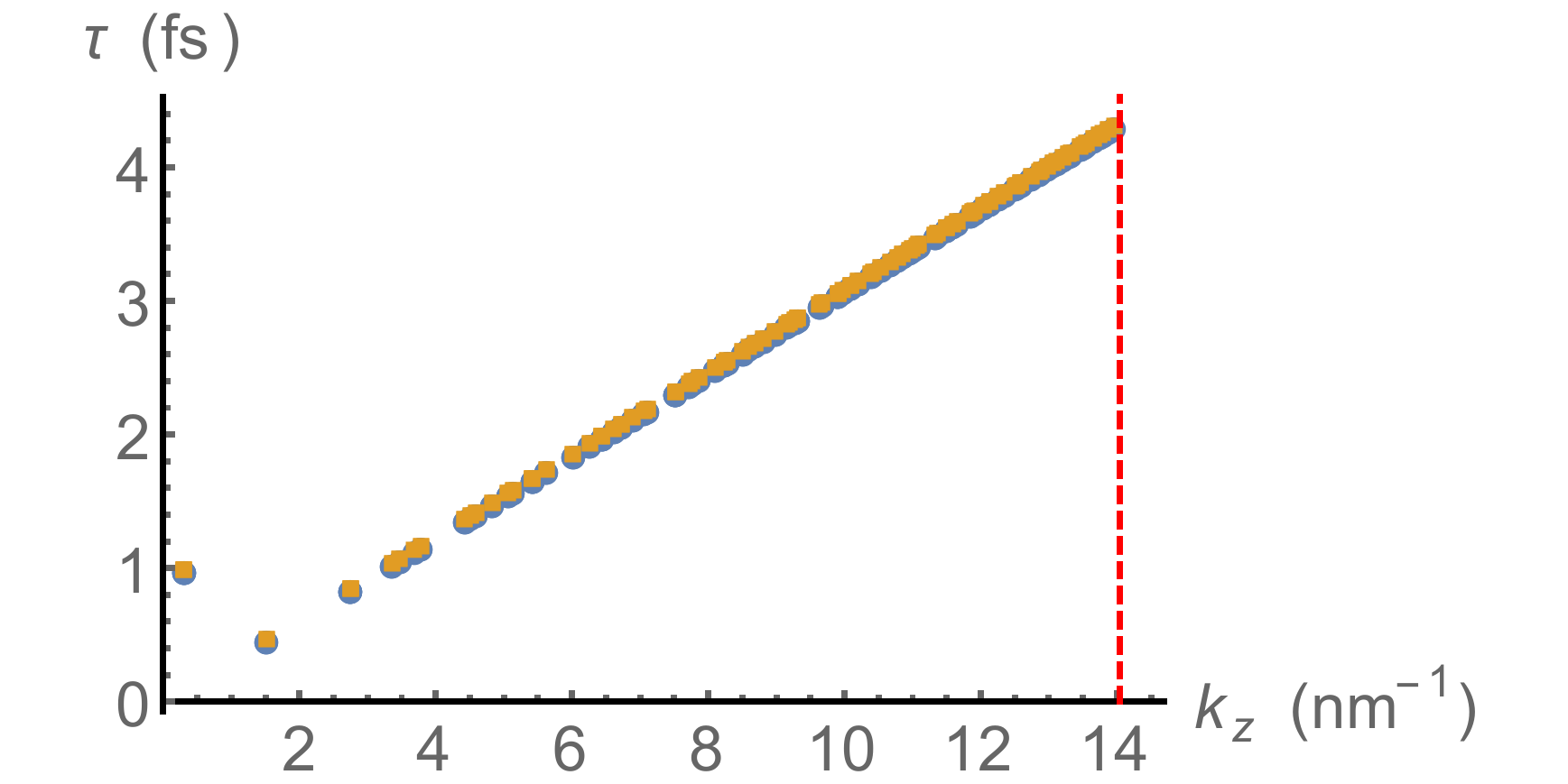}}
\subfigure[]{\includegraphics[width=0.39\linewidth]{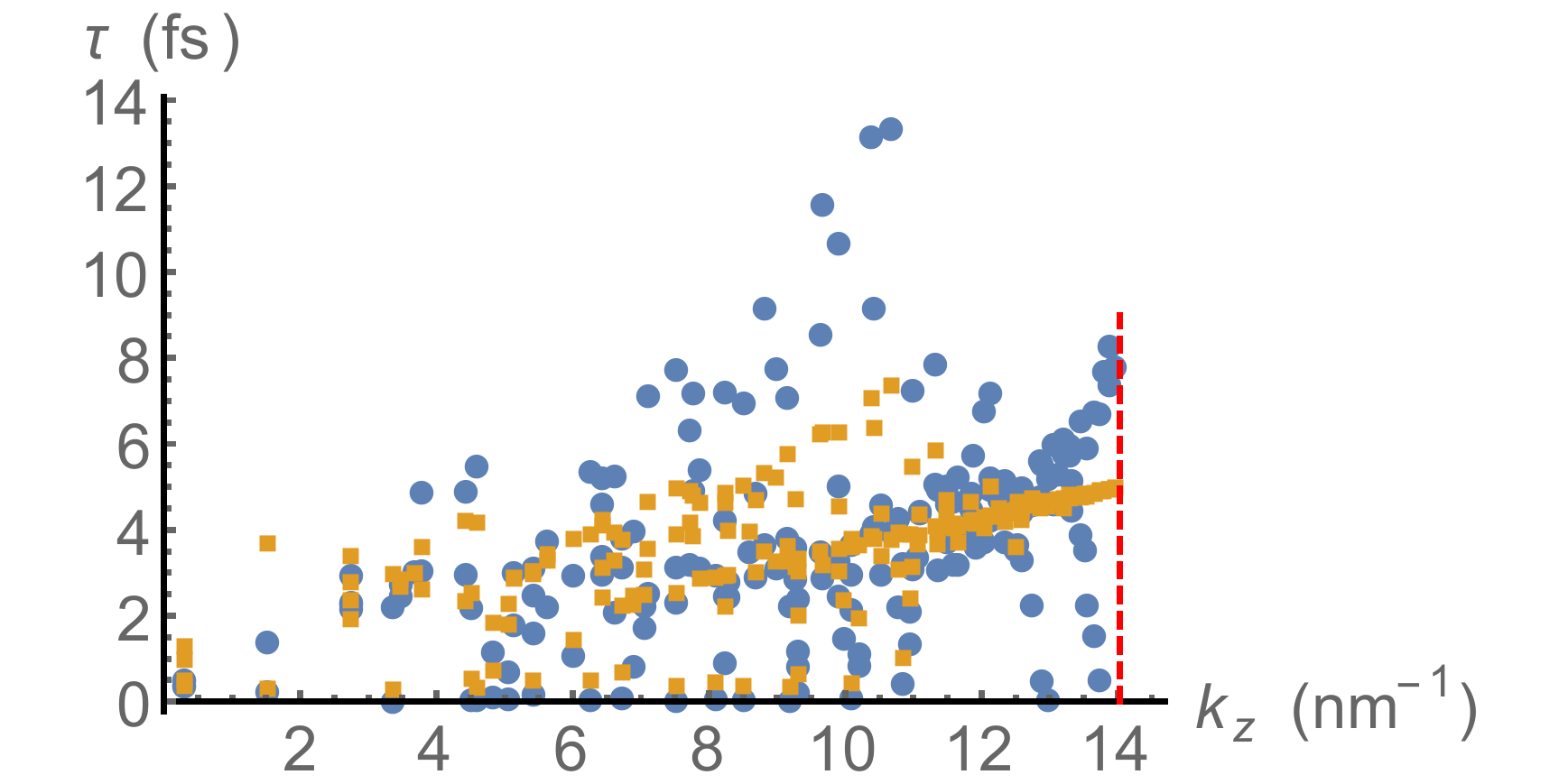}}
\caption{\label{RTPlot} The relaxation times (correct: blue circles, assuming equal RT: yellow squares) for all Fermi level states of a wire with cross-section of 10 by 10 $a_\textnormal{Cu}$ ($D \approx 3.6$ nm) are shown for (a) perpendicular GB planes ($\theta=90^{\circ}$) and (b) tilted GB planes ($\theta=70^{\circ}$) as a function of their transport momentum $k_z$, limited to positive momenta. The constant Fermi wave vector $k_\textnormal{F}$ is shown as a red, dashed line. The simulated wire has $N$ grain boundaries with energy barrier of $S=1.5$eV$\cdot a_\textnormal{Cu}$ and $L_z/N = D$.}
\end{figure}

The first order deviation from the Fermi-dirac equilibrium distribution using (\ref{BTE}), with zero temperature assumed, is given by:
\begin{align*}
f^{(1)}_{n_x n_y}\lef k_z \rig &= -eE_z\tau_{n_x n_y}\lef k_z \rig \frac{\hbar k_z}{m_e} \delta\left[ E_{n_x n_y} \lef k_z \rig - E_\textnormal{F} \right] = -\frac{eE_z}{\hbar} \sum_{\pm} \pm \tau_{n_x n_y}^\pm \delta\lef k_z - k_{z,n_x n_y}^\pm \rig,
\end{align*}
with $k_{z,n_x n_y}^\pm \equiv \pm \sqrt{k_\textnormal{F}^2 - \lef \pi n_x / D \rig^2 - \lef \pi n_y / D \rig^2}$ the positive and negative wave vectors at Fermi level for each sub-band and $\tau_{n_x n_y}^\pm$ the corresponding RTs.
When the scattering probabilities are determined by Fermi's golden rule, the RTs are coupled and have to satisfy the following equations:
\begin{align}
\label{coupledEquations}
1 &= \frac{m_e L_z}{\hbar^3}\sum_{\stackrel{\left\{ n_x', n_y', \pm' \right\}}{\neq \left\{ n_x, n_y, \pm \right\}}} \lef \tau_{n_x n_y}^{\pm} - \frac{k_{z,n_x' n_y'}^{\pm'}}{k_{z,n_x n_y}^\pm} \tau_{n_x' n_y'}^{\pm'} \rig \frac{M_{n_x n_y \pm}^{n_x' n_y' \pm'}}{\left| k_{z,n_x' n_y'}^{\pm'} \right|}, \\ \notag
& M_{n_x n_y \pm}^{n_x' n_y' \pm'} \equiv \left| \langle n_x n_y \pm \mid V \mid n_x' n_y' \pm' \rangle \right|^2,
\end{align}
with $V$ the potential representing the deviations of the nanowire from the ideally conducting box. The sub-band state with reverse wave vector $-k_z$ has the same RT as the state with $k_z$ due to reflection symmetry of the equations. This implies that when only perpendicular GBs are considered, which limits scattering to the Umklapp process,  that the equations for different RTs decouple. Assuming that the RTs are equal when solving (\ref{coupledEquations}), an assumption that one often uses in simulations (e.g. in \cite{jin2007modeling}), gives the correct result (see Fig.~\ref{RTPlot} (a)). However, when tilted GB planes are considered, other scattering events are allowed and the equal RT assumption does not hold (see Fig.~\ref{RTPlot} (b)).

The RT solutions of these equations determine the conductivity $\sigma$ and resistivity $\rho = 1/\sigma$ through:
\begin{align}
\sigma = \frac{e^2}{\pi m_e D^2}\sum_{n_x, n_y, \pm} \tau^{\pm}_{n_x n_y} \left| k_{z,n_x n_y}^{\pm} \right|.
\label{RTAconductivity}
\end{align}
It is clear that any way of approximating the sub-band dependent RT solutions has a direct impact on the resulting resistivity and its scaling.

In what follows, we only consider GBs to be present in $V=V_\textnormal{GB}$, while being modeled as Dirac delta barrier planes distributed along the wire: 
\begin{align}
\label{GBPotential}
V_\textnormal{GB}(x,y,z) = \sum_{\alpha} U_\textnormal{GB} L_\textnormal{GB} \delta \lef z - z_\alpha(x,y) \rig,
\end{align}
with $U_\textnormal{GB}$, $L_\textnormal{GB}$ determining the scattering strength, representing the energy barrier and width of the GB respectively, while the functions $z_\alpha(x,y)$ determine the positions of the GBs.

In \cite{mayadas1970electrical}, there are $N$ perpendicularly oriented GB planes distributed along the wire with the following distribution:
\begin{align*}
g\lef z_1, \ldots, z_N \rig = \frac{\exp\left[-\sum\limits_{\alpha = 1}^{N-1} \lef z_{\alpha+1} - z_\alpha - \frac{L_z}{N} \rig^2/2\sigma_\textnormal{\tiny GB}^2\right]}{L_z \lef 2\pi \sigma_\textnormal{\tiny GB}^2 \rig^{(N-1)/2}},
\end{align*}
with $\sigma_\textnormal{GB}$ the standard deviation of a GB from the uniform distribution along the wire. This distribution yields the following average value for the squared absolute value of the matrix element, $M_{n_x n_y \pm}^{n_x' n_y' \pm'}$:
\begin{align*}
& \left< \left| \langle n_x n_y \pm \mid V_\textnormal{GB} \mid n_x n_y \mp \rangle \right|^2 \right>_{z_\alpha} \approx \frac{N}{L_z^2} \frac{ U_\textnormal{GB}^2 L_\textnormal{GB}^2 \sinh\left[ 2 \lef k_{z, n_x n_y}^\pm \sigma_\textnormal{\tiny GB} \rig^2 \right]}{\cosh\left[ 2 \lef k_{z, n_x n_y}^\pm \sigma_\textnormal{\tiny GB} \rig^2 \right] - \cos \left[ 2k_{z, n_x n_y}^\pm L_z/N \right]}.
\end{align*}
Furthermore we have considered two extensions of this model to introduce some tilt of the GBs, either having a fixed tilt angle $\theta$ in \ref{sectionFixedAngle}, or being randomly tilted in \ref{sectionRandomAngle}.

\setlength{\unitlength}{1.00cm}
\begin{figure}[htb]
\begin{center}
\subfigure[]{\includegraphics[width=0.39\linewidth]{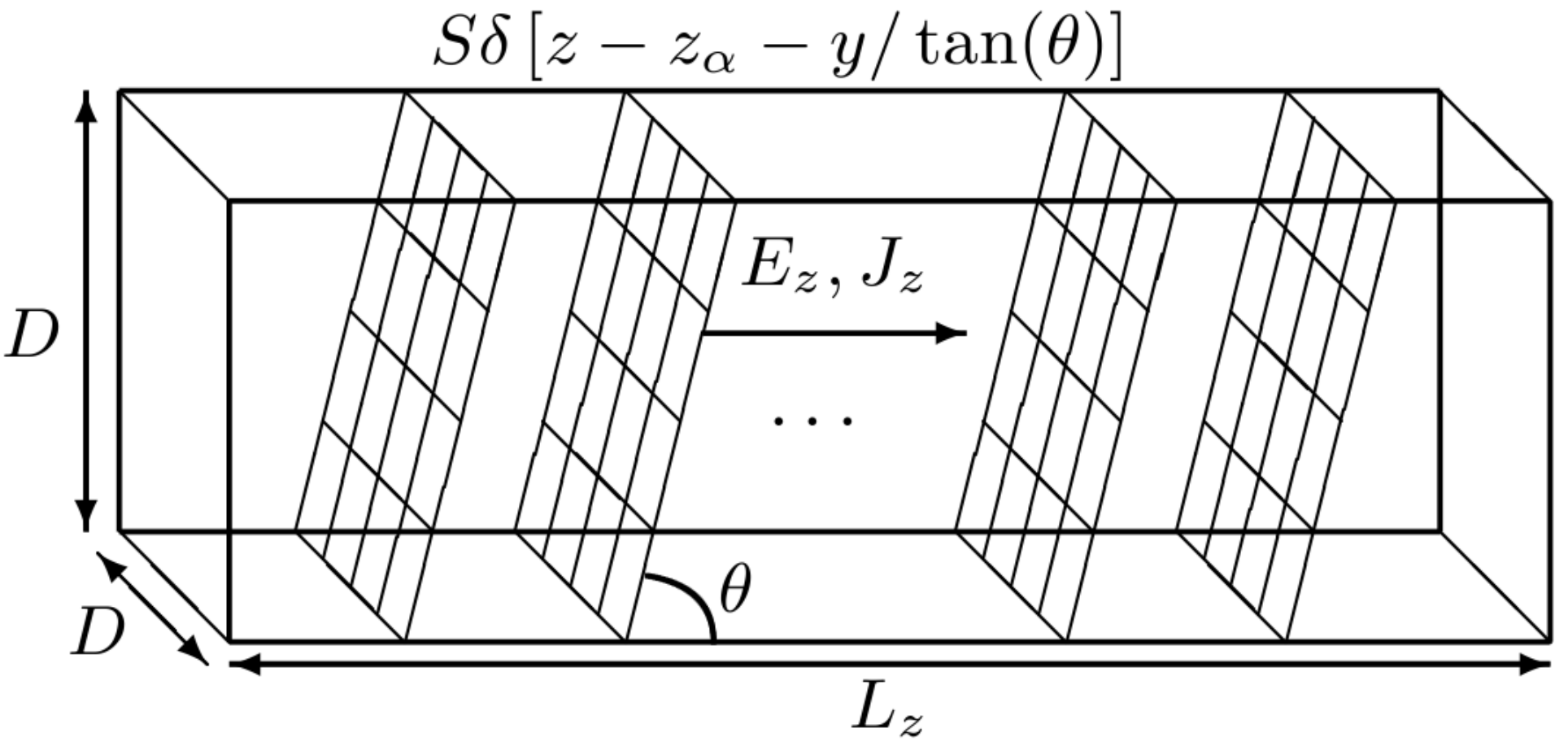}}
\subfigure[]{\includegraphics[width=0.39\linewidth]{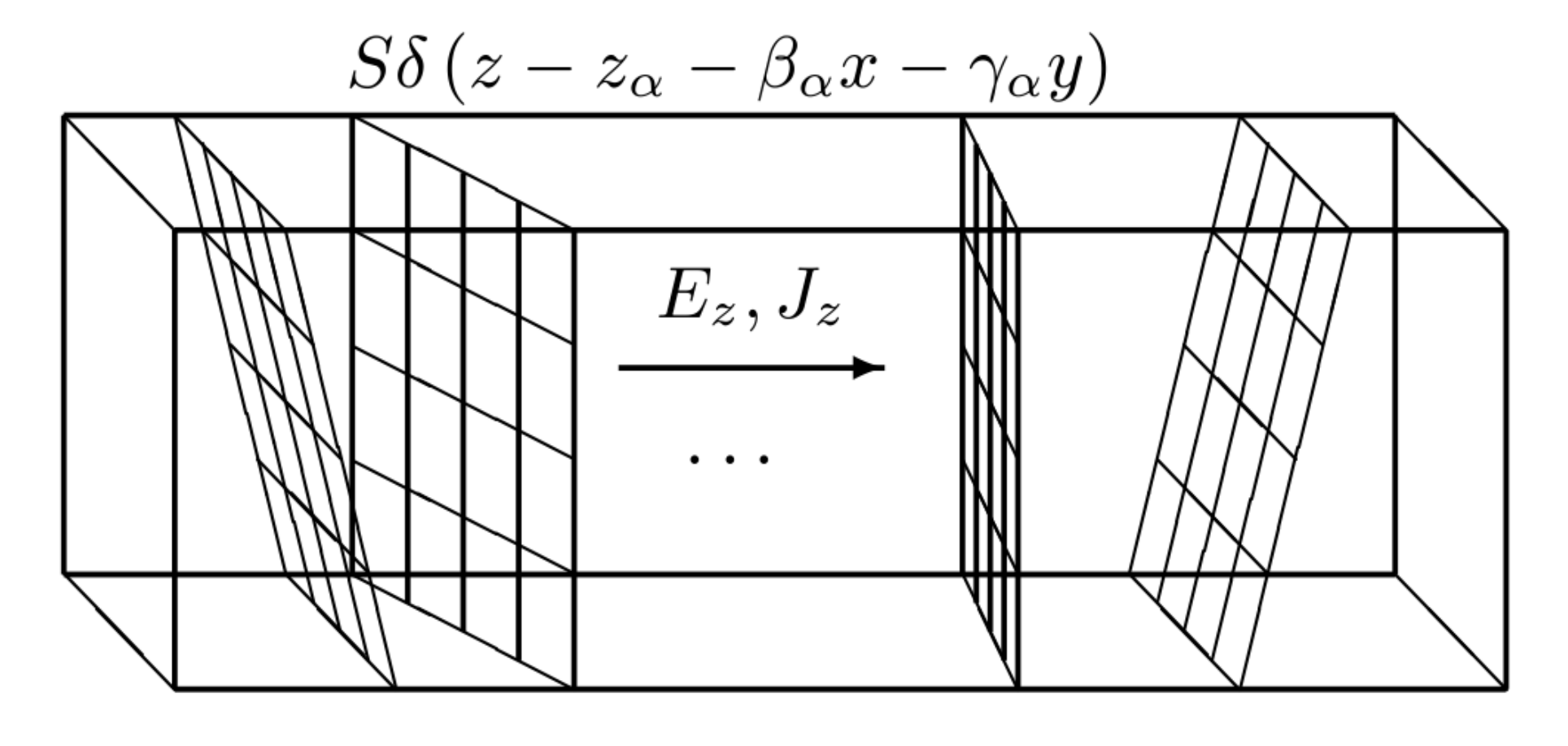}}
\end{center}
\caption{(a) The grain boundaries are all tilted w.r.t. the $y$-axis with the same angle $\theta$. (b) The grain boundaries are randomly tilted, determined by the plane parameters $\beta_\alpha$, $\gamma_\alpha$, uniformly distributed in the interval $\left[ -\Delta, +\Delta \right]$.}
\label{wireModel}
\end{figure}

\begin{figure}[htb]
\centering
\subfigure[]{\includegraphics[width=0.39\linewidth]{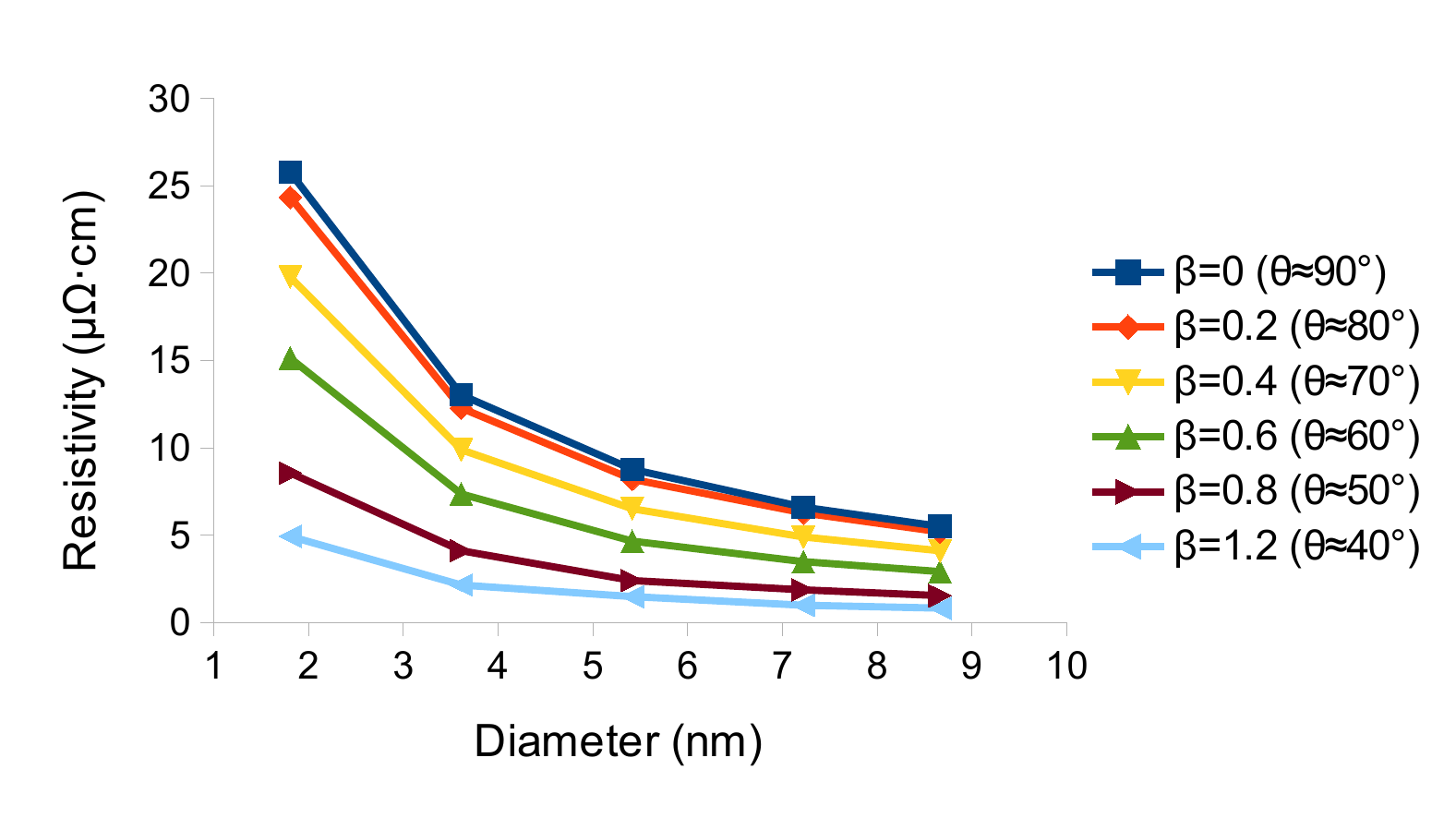}}
\subfigure[]{\includegraphics[width=0.39\linewidth]{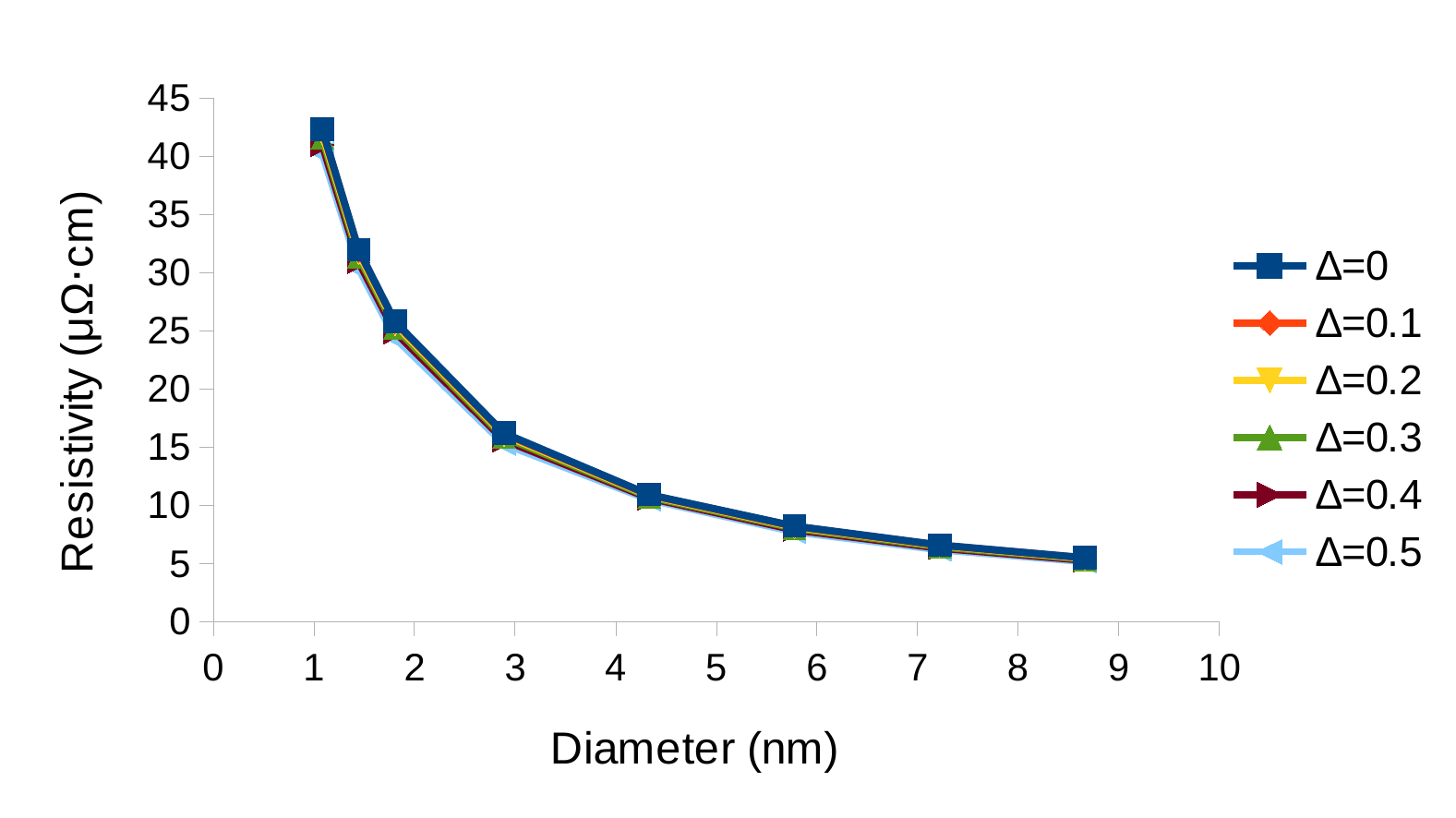}}
\caption{\label{ResPlot} The resistivity of simulated wires with grain boundaries planes tilted (a) with a fixed angle $\theta$ and (b) randomly with a tilt parameter $\Delta$ are shown as a function of the wire diameter. The grain boundaries have energy barrier $S = 1.5$eV$ \cdot a_\textnormal{Cu}$ and $L_z/N = D$.}
\end{figure}

\subsection{GB planes tilted under fixed angle}
\label{sectionFixedAngle}
We consider $N$ GBs with the same distribution as in the MS model, but with all the planes tilted under the same angle, the potential can be modeled by two extra parameters compared to the MS model:
\begin{align*}
V_\textnormal{GB}(x,y,z) = \sum_{\alpha=1}^N U_\textnormal{GB} L_\textnormal{GB} \delta \lef z - z_\alpha^0 - \beta_0 x - \gamma_0 y \rig.
\end{align*}
The squared matrix element changes in the following way:
\begin{align*}
&\left< \left| \langle n_x n_y \pm \mid V_\textnormal{GB} \mid n_x' n_y' \pm' \rangle \right|^2 \right>_{z^0, \beta, \gamma} \\
& \quad \approx \frac{N U_\textnormal{GB}^2 L_\textnormal{GB}^2}{L_z^2} \frac{\sinh\left[ \frac{1}{2} \lef \Delta k \sigma_\textnormal{\tiny GB} \rig^2 \right]}{\cosh\left[ \frac{1}{2} \lef \Delta k \sigma_\textnormal{\tiny GB} \rig^2 \right] - \cos \left[ \Delta k L_z/N \right]} \\
& \quad \quad \times \frac{4}{L_x^2} \left| \int\limits_0^{L_x} \deriv x \, \sin\lef \frac{ n_x \pi x }{ L_x } \rig \sin\lef \frac{ n_x' \pi x }{ L_x } \rig e^{-i\Delta k \beta^0 x} \right|^2 \times \frac{4}{L_y^2} \left| \int\limits_0^{L_y} \deriv y \, \sin\lef \frac{ n_y \pi y}{ L_y } \rig \sin\lef \frac{ n_y' \pi y }{ L_y } \rig e^{-i\Delta k \gamma^0 y} \right|^2
\end{align*}
The only approximations that were made are the same as in the MS model, namely that $N$ is quite large and $\sigma_\textnormal{\tiny GB} \ll L_z$.

In Fig.~\ref{ResPlot}, simulation results for resistivities of different diameter wires are shown for GB planes tilted in the $x$-direction with several values of the tilt angle. The tilt angle $\theta$ is in this case related to the $\beta_0$ by $\tan\theta = \beta_0$.

\subsection{Randomly tilted GB planes}
\label{sectionRandomAngle}
If every GB plane has a different orientation, we would need two parameters per plane to describe the potential. For the sake of simplicity however, we consider an ensemble of random tilts $V_\alpha(x,y,z) = U_\textnormal{GB} L_\textnormal{GB} \delta \lef z-z_\alpha^0 - \beta_\alpha x - \gamma_\alpha y \rig$, with uniformly distributed $\beta_\alpha,\gamma_\alpha \in \left[ -\Delta_{x/y}/2, \Delta_{x/y}/2 \right]$. The two remaining parameters that represent the GB plane tilt in $x$- and $y$-direction respectively are $\Delta_x$ and $\Delta_y$. Hence the tilt distribution is considered to be the same for each GB plane and uniform. The advantage of this distribution, as compared to a uniform distribution in angular coordinates, is that the scattering matrix elements can still be obtained analytically, while capturing a random tilt of each GB. With this ensemble, one has to be careful for large values of $\Delta_{x/y}$ however. The distribution of tilt angles will be dominated by large angles because of the Jacobian when transforming to spherical coordinates.

The constants $\Delta_{x/y}$ define a maximum tilt for the GBs. The squared matrix element averaged over GB positions and orientations reads:
\begin{align*}
& \left< \left| \langle n_x n_y \pm \mid V_\textnormal{GB} \mid n_x' n_y' \pm' \rangle \right|^2 \right>_{z^0, \beta, \gamma} \\
& \quad \approx \frac{N U_\textnormal{GB}^2 L_\textnormal{GB}^2}{L_z^2} \frac{\sinh\left[ \frac{1}{2} \lef \Delta k \sigma_\textnormal{\tiny GB} \rig^2 \right]}{\cosh\left[ \frac{1}{2} \lef \Delta k \sigma_\textnormal{\tiny GB} \rig^2 \right] - \cos \left[ \Delta k L_z/N \right]} \\
%& \quad \quad \times \frac{4}{L_x^2} \left| \int\limits_0^{L_x} \deriv x \, \frac{2  \sin\lef \frac{ n_x \pi x}{L_x} \rig \sin\lef \frac{n_x' \pi x}{L_x} \rig \sin \left[ \frac{\Delta_x \, \Delta k \, x}{2} \right]}{\Delta k \, \Delta_x x} \right|^2 \quad \times \frac{4}{L_y^2} \left| \int\limits_0^{L_y} \deriv y \, \frac{2 \sin\lef \frac{n_y \pi y}{ L_y } \rig \sin\lef \frac{ n_y' \pi y }{ L_y } \rig \sin \left[ \frac{\Delta_y \, \Delta k \, y}{2} \right]}{ \Delta k \, \Delta_y y }\right|^2
& \quad \quad \times \frac{4}{L_x^2} \int\limits_0^{L_x} \deriv x \int\limits_0^{L_x} \deriv x' \, \sin\lef \frac{ n_x \pi x}{L_x} \rig \sin\lef \frac{n_x' \pi x}{L_x} \rig \sin\lef \frac{ n_x \pi x'}{L_x} \rig \sin\lef \frac{n_x' \pi x'}{L_x} \rig \frac{2 \sin \left[ \frac{\Delta_x \, \Delta k \, \lef x - x' \rig}{2} \right]}{\Delta k \, \Delta_x \lef x - x' \rig} \\
& \quad \quad \times \frac{4}{L_y^2} \int\limits_0^{L_y} \deriv y \int\limits_0^{L_y} \deriv y' \, \sin\lef \frac{n_y \pi y}{ L_y } \rig \sin\lef \frac{ n_y' \pi y }{ L_y } \rig \sin\lef \frac{n_y \pi y'}{ L_y } \rig \sin\lef \frac{ n_y' \pi y' }{ L_y } \rig \frac{2 \sin \left[ \frac{\Delta_y \, \Delta k \, \lef y - y' \rig}{2} \right]}{ \Delta k \, \Delta_y \lef y - y' \rig },
\end{align*}
with $\Delta k \equiv \lef k_{z,n_x n_y}^\pm - k_{z,n_x' n_y'}^{\pm'} \rig$.

The resistivities obtained with simulations of randomly oriented GB planes for different values of tilt parameter $\Delta \equiv \Delta_x = \Delta_y$ are shown in Fig.~\ref{ResPlot} (b).

\section{Conclusion and remarks}
We have simulated the resistivity for metal nanowires with cross sections up to 10nm by 10nm with GB planes tilted under a fixed angle or randomly tilted as a scattering source for the electrons. The resistivity values clearly decrease for all diameters when the GB planes are randomly tilted, as shown in Fig.~\ref{ResPlot}. When each GB plane is randomly tilted, there is also a lower resistivity w.r.t. perpendicularly oriented GB planes, but the difference is barely visible. An important aspect of these resistivity values is the correct retrieval of the sub-band state dependent RTs. From Fig.~\ref{RTPlot} it is apparent that there is no clear relation between the transport momentum and the RT of an electron and that approximate methods to obtain the RTs can give incorrect results. It is very important however to have precise results for this relation, because it has a direct impact on the resistivity and conductivity through (\ref{RTAconductivity}). In the case of a fixed tilt angle for all GB planes, the approximation assuming equal RTs in (\ref{coupledEquations}), shown in Fig.~\ref{RTPlot} (b), underestimates the conductivity, but the reverse effect, an overestimation, can also be observed if other scattering mechanisms are considered \cite{moors2014resistivity}. In the case of tilted GB planes, a lot of very stable outliers and a stability trend for high transport momentum electron states are missed by the equal RT approximation, neither of these effects being present in case of perpendicular GB planes. Very interestingly, the RTs for high transport momentum electron states goes up and approaches the value obtained with another approximation method, ignoring the incoming scattering while applying Fermi's golden rule.

The two profiles of GB tilt that were presented in this text ignore some aspects of GBs. As a first remark, the resistivities shown in Fig.~\ref{ResPlot} (a) only show GB planes tilted in the $x$-direction. When the GBs are tilted in a combination of the $x$- and $y$-direction, the results will be quantitatively different, but the overall trend remains the same. A second remark addresses the correlations between the orientations of two different GB planes. For low GB densities, it is a safe approximation to neglect these correlations as we have done, but for higher densities correlations between adjacent GB planes can be expected. Deviations from the potential form in (\ref{GBPotential}) can as well be modeled, definitely for smaller diameters. Further input of GB properties and statistics in small diameter metal nanowires is required to make more realistic GB models. Other correlations that could be included are those between the tilt angle and the energy barrier height and/or width. When those correlations are observed experimentally, it would be interesting to include them in the statistical ensemble of $V_\textnormal{GB}$. If the energy barrier height or width would increase substantially for tilted GBs, the decrease in resistivity for larger tilt angles could be reduced. However, it is expected that a more realistic description of GBs in metal nanowires would still show a resistivity decrease for increasing tilt angles. It is easy to show that in the limit where all GBs are tilted under $90^\circ$, we retrieve a form of surface roughness scattering as described by Ando's model, which has a much lower resistivity contribution \cite{moors2014resistivity}.

\section*{Acknowledgment}
We would like to thank Christian Maes for many useful discussions.

\bibliographystyle{IEEEtran}
\bibliography{IEEEabrv,confPaper}{}

\end{document}